\documentclass[11pt]{article}
\usepackage{amsfonts}
\usepackage{mathrsfs}
\usepackage{graphicx}
\usepackage{amsmath}
\usepackage{amssymb}
\usepackage{subfigure}
\usepackage{epsfig}
\usepackage{graphicx}
\usepackage{color}
\usepackage{indentfirst}
\usepackage{hyperref}
\usepackage{setspace}
\usepackage{braket}
\usepackage{slashed}
\usepackage{cite}
\usepackage{multirow}
\usepackage{cancel}
\usepackage{caption}
\usepackage{array}
\usepackage{appendix}
\usepackage{tabularx}
\usepackage{tabulary}

\hypersetup{colorlinks=true, citecolor=red, urlcolor=blue, linkcolor=blue}
\begin{document}
\title{Hyperon semileptonic decays in QCD sum rules}
\author{Sheng-Qi Zhang$ ^{1} $, Xuan-Heng Zhang$ ^{1} $ and Cong-Feng Qiao$^{1}\footnote{qiaocf@ucas.ac.cn.}$\\
\\
\normalsize{$^{1}$School of Physical Sciences, University of Chinese Academy of Sciences}\\
\normalsize{YuQuan Road 19A, Beijing 100049, China}\\
\\
}
\date{}
\maketitle

\begin{abstract}
We investigate the hyperon semileptonic decays within the framework of QCD sum rules. The flavor $ SU(3) $ symmetry breaking effects are analyzed via the relevant form factors and corresponding branching fractions. Employing the $ z $-series parameterization to capture the $ q^2 $ dependence of form factors, we calculate the hyperon semileptonic decay rates and confront them with the recent experimental measurements. Moreover, we calculate as well the non-standard tensor form factors, which involve certain new physics beyond the standard model.
\vspace{0.3cm}
\end{abstract}

\section{Introduction}
Hyperon semileptonic decays (HSDs) play an important role in the exploration of strong and weak interactions in light baryon sector, where the former involves the hadronization of partons, and the latter associate with the flavor-changing processes. The precise measurement of the Cabibbo-Kobayashi-Maskawa (CKM) matrix element $ | V_{us} |$ can offer substantial test for the standard model (SM) and potential evidence for new physics beyond the SM. Furthermore, it is anticipated that the decays of light baryons with strange quarks (hyperons) may stand as potential probes for $CP$ violation as well \cite{Donoghue:1985ww, Fu:2023ose}.

HSDs were ever described by the Cabibbo model featuring the exact $ SU(3) $ symmetry~\cite{Cabibbo:1963yz}. However, the experimental data for HSDs suggest the potential breaking of flavor $SU(3)$ symmetry. For example, processes $ n\rightarrow p e \bar{\nu}_e $ and $ \Xi^0\rightarrow \Sigma^+ e \bar{\nu}_e $ find that the ratio of the axial-vector over vector form factors $g_1(0)/f_1(0) $ are  $ 1.2754\pm 0.0013 $ and $ 1.22\pm 0.05 $~\cite{ParticleDataGroup:2022pth, KTeV:2001djr, NA48I:2006yat}, respectively. The difference of these results implies certain deviation from the $SU(3)$ symmetry since they should be equal in exact $ SU(3) $ symmetry. More comprehensive understanding of flavor $ SU(3) $ symmetry breaking necessitates more precise experimental measurements. However, the measurement of HSD process remains to be an experimental challenge due to its tiny branching fractions~\cite{ParticleDataGroup:2022pth} and complexity in distinguishing semileptonic decays from dominant two-body decay backgrounds~\cite{Cabibbo:2003cu}. This underscores the considerable scope for the enhancement on both the experimental and theoretical fronts. Hopefully, the proposed super $ \tau$-charm facility (STCF) could offer a distinctive opportunity for hyperon studies, leveraging high-statistics samples of hyperon-antihyperon pair events and time-like form factor measurements~\cite{Achasov:2023gey}.

Moreover, HSDs offer an exceptional opportunity for extracting the CKM matrix element $ | V_{us} | $, while the current determination of $ |V_{us}| $ heavily depends on kaon decays~\cite{ParticleDataGroup:2022pth}. In recently, BESIII collaboration reported a new result of $ \Lambda\rightarrow p\mu^-\bar{\nu}_\mu  $~\cite{BESIII:2021ynj}. The absolute branching fraction is determined to be $ \mathcal{B}(\Lambda\to p\mu^-\bar{\nu}_\mu)=(1.48\pm 0.21\pm 0.08)\times10^{-4} $. This result along with earlier ones provide an opportunity to extract a precise $| V_{us} |$, playing an ancillary role to test the quark-flavor mixing in the hyperon sector. Theoretically, a model-independent assessment of $SU(3)$ breaking effects is pivotal for the extraction of $ | V_{us} | $ \cite{Gaillard:1984ny, Cabibbo:2003cu}, although it appears that the HSDs data still align well with the unbroken $SU(3)$ predictions due to the smallness of hyperons mass splitting ($\sim$$15\%$).

Up to now, various approaches have been employed to the analysis of $ SU(3) $ breaking effects of HSDs, including different quark models~\cite{Donoghue:1986th, Schlumpf:1994fb, Faessler:2008ix, Migura:2006en}, soliton model~\cite{Kim:1997ts, Kim:1999uf, Yang:2015era, Ledwig:2008ku}, chiral perturbation theory ($ \chi $PT)~\cite{Kaiser:2001yc, Villadoro:2006nj, Ledwig:2014rfa, Geng:2009ik}, $ 1/N_c $ and large $ N_c $ expansion~\cite{Flores-Mendieta:1998tfv, Flores-Mendieta:2004cyh}, $ SU(3) $ Skyrme model~\cite{Park:1989nz, Kondo:1991fc}, and lattice QCD~\cite{Cooke:2012xv, Cooke:2013qqa, Sasaki:2008ha, Sasaki:2012ne, Guadagnoli:2006gj, Sasaki:2017jue}. Additionally, the technique of three-point QCD sum rules (QCDSR), initially introduced by Ioffe \textit{et al}. to investigate the pion electromagnetic form factors~\cite{Ioffe:1982ia, Nesterenko:1982gc}, has been extensively employed to the study of baryon semileptonic decays~\cite{Dai:1996xv, Zhang:2023nxl, Huang:1998rq, Huang:1998ek, MarquesdeCarvalho:1999bqs, Shi:2019hbf, Zhao:2020mod, Zhao:2021sje, Xing:2021enr}. In contrast to a phenomenological model, QCDSR is a theoretical framework based on the QCD first principle. QCDSR starts by formulating three-point correlation functions through appropriate interpolating current operators, where nonperturbative effects are parameterized as vacuum expectation values at each dimension. Once the equivalence between the QCD and phenomenological representation of the three-point correlation functions is established via quark-hadron duality, the form factors arising in weak transitions are formally ascertained. The foundational physical assumptions of QCDSR are more closely associated with field theory~\cite{Ball:1991bs}, a connection that may align more effectively with the argument of "model-independence" in Ref.~\cite{Cabibbo:2003cu}. In this study, we employ QCDSR to evaluate the transition form factors of HSDs, afterwards the $ SU(3) $ breaking effects and the branching fractions of HSDs can be analyzed. Moreover, we also present the results of the non-standard tensor form factors, which may stand as probes for the new physics beyond the SM \cite{Chang:2014iba}.

The rest of the paper is organized as follows: in Sec.~\ref{Formalism} we interpret the fundamental concept of QCDSR for the three-point correlation functions. The numerical results and analysis are detailed in Sec.~\ref{Numerical}. The conclusions and discussions are presented in the final section.

\section{Formalism}\label{Formalism}

There are a total six transition modes of HSDs, which are $\Lambda\rightarrow p, \, \Sigma^0\rightarrow p, \, \Xi^-\rightarrow \Lambda, \, \Xi^-\rightarrow \Sigma^0, \, \Xi^0\rightarrow \Sigma^+$, and $\Sigma^-\rightarrow n$. One is only bothered to consider the first four of them, and the other two may be determined through isospin relations \cite{Yang:2015era, Ledwig:2008ku}. The transition matrix element for the generic HSDs process can be formulated as
\begin{equation}
\mathcal{M}_{B_1\rightarrow B_2 \ell\nu_\ell} = \frac{G_F}{\sqrt{2}}V_{u s}[\bar{\ell}\gamma^\mu\left(1-\gamma_5\right)\nu_\ell]\langle B_2(q_2)|\bar{u}\gamma_\mu(1-\gamma_5)s|B_1(q_1)\rangle \; ,
\end{equation}
where $ B_1 $ and $ B_2 $ are the initial and final state baryons. $ G_F $ represents the Fermi constant, and $ V_{u s} $ is the CKM matrix element. On the other hand, the hadronic matrix element can be described in terms of transition form factors:
\begin{align}
\label{matrix}
\bra{B_2(q_2)}\bar{u}\gamma_\mu\left(1-\gamma_5\right)s\ket{B_1(q_1)}&=\bar{u}_{2}(q_2) \bigg[f_1(q^2) \gamma_\mu+i f_2(q^2)\sigma_{\mu\nu}\frac{q^\nu}{M_{1}}+f_3(q^2) \frac{q_\mu}{M_{1}} \bigg] u_1(q_1)\nonumber\\
&-\bar{u}_{2}(q_2)\bigg[g_1(q^2) \gamma_\mu+i g_2(q^2) \sigma_{\mu\nu}\frac{q^\nu}{M_{1}}+g_3(q^2) \frac{q_\mu}{M_{1}}\bigg]\gamma_5 u_1(q_1)\;.
\end{align}
Here, $ q_1 $ and $ q_2 $ are the four-vector momentum of the initial and final state baryons, respectively. The momentum transfer $ q=q_1-q_2 $ and $ M_1 $ is the mass of the initial state baryon. These real-valued form factors are functions of Lorentz invariant $ q^2 $, due to the requirement of time reversal invariance \cite{Gaillard:1984ny, Schlumpf:1994fb}. They are known as the vector and axial-vector form factors $ f_1 $ and $ g_1 $, the weak magnetism and electric form factors $ f_2 $ and $ g_2 $, and the induced scalar and pseudoscalar form factors $ f_3 $ and $ g_3 $. The weak electric form factor $ g_2 $, classified as the second-class form factor by Weinberg~\cite{Weinberg:1958ut}, is often assumed to be very small~\cite{Cabibbo:2003ea, Schlumpf:1994fb, Yang:2015era}. The induced scalar and pseudoscalar form factors $ f_3 $ and $ g_3 $ can be safely neglected in the calculation since their contributions to hyperon semileptonic decays are suppressed significantly by a factor of $ (m_{\ell}/M_1)^2 $~\cite{Schlumpf:1994fb, Sasaki:2008ha}, where $ m_\ell( \ell = e, \mu ) $ denote the lepton masses.

Through the above analysis, and considering the relatively more abundant experimental data on $ f_1 $, $ f_2 $, and $ g_1 $, we will focus merely on these three form factors. Note that in the Cabibbo model, $f_1(0)$ is well determined, $f_2(0)$ can be determined in terms of anomalous magnetic moments of proton and neutron, and $g_1(0)$ appears to be a function of $F$ and $D$ \cite{Gaillard:1984ny}:
\begin{align}
\label{g10}
g_1(0)&=a\cdot F+b\cdot D\;,
\end{align}
where $ a $ and $ b $ are generalised Clebsch-Gordan coefficients of $SU(3)$ group. Here, $F$ and $D$ denote the coupling constants, resulting from Cabibbo's fundamental assumption that the weak hadronic currents belong to a single self-conjugate representation of $SU(3)$. The Cabibbo model predictions for these three form factors are given in Table~\ref{table:SU3}.
\begin{table}[ht]
\centering
\caption{Form factors of HSDs predicted by the Cabibbo model. $ \mu_p $ and $ \mu_n $ stand for the anomalous magnetic moments of the proton and neutron, respectively.}
\begin{tabular}{lcccc}
\hline\hline
& $f_1(0)$ & $g_1(0)$ & $g_1(0)/f_1(0)$ & $f_2(0)/f_1(0)$  \\
\hline
$\Lambda\rightarrow p$ & $-\sqrt{\frac{3}{2}}$& $-\sqrt{\frac{3}{2}}(F+\frac{D}{3})$ &$F+\frac{D}{3}$& $\frac{M_\Lambda}{M_p}\frac{\mu_p}2$\\
$\Sigma^{0}\rightarrow p$ & $-\sqrt{\frac{1}{2}}$& $-\sqrt{\frac{1}{2}}(F-D)$ &$F-D$& $\frac{M_{\Sigma^0}}{M_p}\frac{(\mu_p+2\mu_n)}2$ \\
$\Xi^{-}\rightarrow \Lambda$ & $\sqrt{\frac{3}{2}}$& $\sqrt{\frac{3}{2}}(F-\frac{D}{3})$ &$F-\frac{D}{3}$& $-\frac{M_{\Xi^-}}{M_p}\frac{(\mu_p+\mu_n)}2$ \\
$\Xi^{-}\rightarrow \Sigma^{0}$ & $\sqrt{\frac{1}{2}}$& $\sqrt{\frac{1}{2}}(F+D)$ &$F+D$& $\frac{M_{\Xi^-}}{M_p}\frac{(\mu_p-\mu_n)}2$ \\
\hline\hline
\end{tabular}
\label{table:SU3}
\end{table}

To calculate the above three form factors in the framework of QCDSR, the three-point correlation functions could be constructed, as
\begin{equation}
\label{3pcf}
\Pi_\mu(q_1^2, q_2^2, q^2)=i^2 \int d^4 x\; d^4 y\; e^{i(-q_1 x+q_2 y)} <0|T\{j_{2}(y)j_\mu(0)j^{\dagger}_{1}(x) \}|0>\;.
\end{equation}
The weak transition current $ j_\mu = \bar{u} \gamma_\mu\left(1-\gamma_5\right) s $. $j_1$ and $j_2$ represent the interpolating currents for the initial and final state baryons, formally defined as:~\cite{Shuryak:1981fza, Huang:1998rq, Huang:1998ek}:
\begin{align}
\label{baryon-current}
j&=\epsilon^{a b c}\left(q^{a T}_i C \sigma^{\mu\nu} q^b_j\right)\sigma_{\mu\nu} \gamma_5 q^c_k\;,
\end{align}
where the superscripts $ a $, $ b $, and $ c $ represent the color indices and $ C $ is the charge conjugation matrix. The subscripts $ i $, $ j $, and $ k $ denote different flavor of light quarks. In our calculation, $ (i, j, k) = (u,u,d), (u,s,d), (u,d,s), $ and $ (s,s,d) $ for $ p,\, \Lambda,\, \Sigma^0, $ and $ \Xi^- $, respectively.

On the QCD side, the three-point correlation functions presented in Eq.~(\ref{3pcf}) can be formulated using operator-product expansion (OPE) and double dispersion relations:
\begin{align}
\label{3ptQCD}
\Pi_\mu^{\text{QCD}}(q_1^2, q_2^2, q^2)=\int_{s_1^\text{min}}^{\infty}d s_1\int_{s_2^\text{min}}^{\infty}d s_2\frac{\rho^{\text{QCD}}_\mu(s_1,s_2,q^2)}{(s_1-q_1^2)(s_2-q_2^2)}\;,
\end{align}
where $ s_{1(2)}^\text{min} $ represents the kinematic limit. The spectral density $ \rho^{\text{QCD}}_\mu(s_1,s_2,q^2) $ can be determined using Cutkosky cutting rules~\cite{Zhao:2020mod, MarquesdeCarvalho:1999bqs,Zhao:2021sje, Shi:2019hbf, Xing:2021enr, Wang:2012hu, Yang:2005bv, Du:2003ja}. In this study, contributions up to dimension 6 are considered in $ \rho^{\text{QCD}}_\mu(s_1,s_2,q^2) $, expressed as:
\begin{align}
\label{spectra-density}
\rho^{\text{QCD}}_\mu(s_1,s_2,q^2)&=\rho^{\text{pert}}_\mu(s_1,s_2,q^2)+\rho^{\langle \bar{q}q\rangle}_\mu(s_1,s_2,q^2)+\rho^{\langle g_s^2G^2 \rangle}_\mu(s_1,s_2,q^2)\nonumber\\
&+\rho^{\langle g_s \bar{q}\sigma \cdot  G q \rangle}_\mu(s_1,s_2,q^2)+\rho^{\langle \bar{q}q \rangle^2}_\mu(s_1,s_2,q^2)\;.
\end{align}
Here the first term corresponds to the perturbative contribution, whereas $ \langle\bar{q}q \rangle$, $  \langle g_s^2G^2 \rangle $, $ \langle g_s \bar{q}\sigma \cdot  G q \rangle $, and $ \langle \bar{q}q \rangle^2 $ characterize nonperturbative effects as different orders of vacuum expected value that are known as the condensates.

On the phenomenological side, by incorporating a complete set of intermediate hadronic states and leveraging double dispersion relations, the three-point correlation functions of Eq.~(\ref{3pcf}) can be expressed as follows:
\begin{align}
\label{spectra}
\Pi_\mu^{\text{phe}}(q_1^2, q_2^2, q^2)=&\sum_{\text{spins}}\frac{\bra{0}j_{2}\ket{B_2(q_2)}\bra{B_2(q_2)}j_\mu\ket{B_1(q_1)}\bra{B_1(q_1)}j_{1}\ket{0}}{(q_1^2-M_{1}^2)(q_2^2-M_{2}^2)}\nonumber\\
+&\text{higher resonances and continuum states}\;,
\end{align}
where $ M_2 $ denotes the mass of the final state baryon, and the transition amplitudes from vacuum to baryon states can be parameterized by introducing the decay constants,
\begin{align}
\bra{0}j_{i}\ket{B_i(p)}&=\lambda_{i}u_{i}(p)\;.
\end{align}
Here, the subscript $ i=1,2 $ represents the notation of initial and final state baryons, respectively. $ p $ is the four-vector momentum carried by the baryons, and $ u_{i} $ denotes the Dirac spinor. $ \lambda_i $ represents the decay constant of baryon octet states, which can be calculated through two-point sum rules~\cite{Ioffe:1981kw, Nesterenko:1983ef, Yang:1993bp, Hwang:1994vp, Zhao:2021sje}. Utilizing the hadronic matrix element presented in Eq.~(\ref{matrix}) and the spin sum completeness relations $ \sum u_{i}(p)\bar{u}_{i}(p)=\slashed{p}+M_{i} $, one can derive the phenomenological representation of the three-point correlation functions of Eq.~(\ref{3pcf}):
\begin{align}
\label{3ptphe}
\Pi_\mu^{\text{phe}}(q_1^2, q_2^2, q^2) & =\frac{\lambda_{2} \left(\slashed{q}_2+M_{2}\right)\bigg[f_1(q^2) \gamma_\mu+i f_2(q^2)\sigma_{\mu\nu}\frac{q^\nu}{M_{1}}+f_3(q^2)\frac{q_\mu}{M_{1}}\bigg]\lambda_{1}\left(\slashed{q}_1+M_{1}\right)}{(q_1^2-M_{1}^2)(q_2^2-M_{2}^2)} \nonumber\\
&-\frac{\lambda_{2} \left(\slashed{q}_2+M_{2}\right)\bigg[g_1(q^2) \gamma_\mu+i g_2(q^2)\sigma_{\mu\nu}\frac{q^\nu}{M_{1}}+g_3(q^2)\frac{q_\mu}{M_{1}} \bigg]\gamma_5\lambda_{1}\left(\slashed{q}_1+M_{1}\right)}{(q_1^2-M_{1}^2)(q_2^2-M_{2}^2)}\nonumber\\
&+\text{higher resonances and continuum states}\;.
\end{align}
It should be noted that despite the presence of twenty-four Lorentz structures in Eq.~(\ref{3ptphe}), only six form factors need to be determined. The redundant Lorentz structures of the form factors are addressed in the contribution of negative-parity baryons~\cite{Jido:1996zw, Jido:1996ia, Zhao:2020mod}, in the "higher resonances and continuum states" term in Eq.~(\ref{3ptphe}).

Adopting quark-hadron duality and double Borel transform, we can establish an equivalence between the two representations in Eqs.~(\ref{3ptQCD}) and (\ref{3ptphe}). Consequently, the general form of the form factors can be obtained:
\begin{align}
\label{form-factors-exp}
f^{B_1\rightarrow B_2}(q^2)=\frac{e^{M_{1}^2/\tau_{1}^2}e^{M_2^2/\tau_{2}^2}}{\lambda_{1}\lambda_2}\int_{s_1^\text{min}}^{s_1^0}d s_1\int_{s_2^\text{min}}^{s_2^0}d s_2\;\rho(s_1,s_2,q^2)e^{-s_1/\tau_1^2} e^{-s_2/\tau_2^2}+\Pi_{0}\;,
\end{align}
where $ s_1^0 $ and $ s_2^0 $ represent the threshold parameters of the initial and final state baryons, respectively. $ \Pi_0 $ comprises contributions in the correlation function devoid of an imaginary part yet manifesting nontrivial features after the Borel transformation. Additionally, $ \tau_{1}^2 $ and $ \tau_{2}^2 $ denote the Borel parameters that emerge after the double Borel transform. For convenience, we showcase the $ f_1(0) $ expression of the transition $ \Lambda\rightarrow p $:
\begin{align}
f_1^{\Lambda\rightarrow p}(0)&=\frac{e^{M_{1}^2/\tau_{1}^2}e^{M_2^2/\tau_{2}^2}}{\lambda_{1}\lambda_2 M_{1}^{\prime}M_{2}^{\prime} }\bigg[\int_{s_1^\text{min}}^{s_1^0}d s_1\int_{s_2^\text{min}}^{s_2^0}d s_2\;\frac{3\,s_2^2\,m_s^2 \left(m_s^2-\delta _s\right)e^{-s_1/\tau_1^2} e^{-s_2/\tau_2^2}}{64\,\pi ^4  \delta _s^5}\nonumber\\[5pt]
&\times\big(4 \,m_s^2\, \delta _s \left(2 M_{1}^{-} M_{2}^{-}-s_2\right)+3 \,m_s^4 \left(s_2-3 M_{1}^{-} M_{2}^{-}\right)+\delta _s^2 \left(M_{1}^{-} M_{2}^{-}+3 \,s_2\right)\big)\nonumber\\[5pt]
&-\frac{2 \,\langle \bar{q}q \rangle^2\, m_s (M_{1}^{-}+5 M_{2}^{-})}{3}e^{-m_s^2/\tau_1^2}\bigg]\;.
\label{exp-f1}
\end{align}
Here, we define $ \delta_s = s_1-s_2$ and $ M^\prime = M^{+}+M^{-} $. $ M^{+} $ and $ M^{-} $ stand for the masses of the corresponding positive and negative parity baryons, where the values are given in the next section. It can be observed from Eq.~(\ref{exp-f1}) that the quark-gluon condensate $ \langle g_s \bar{q}\sigma \cdot  G q \rangle $ vanishes in the transition form factors, while the contribution of gluon condensate $ \langle g_s^2G^2 \rangle $ is greatly suppressed by the strong coupling constant in practice.

\section{Numerical results and discussions}\label{Numerical}

\begin{table}[ht]
\centering
\caption{Input parameters of baryon octet.}
\begin{tabular}{ccccccccc}
\hline\hline
& $ p $  & $ \Lambda $  & $ \Sigma^0 $ & $ \Xi^- $ \\ \hline
$ M(\frac{1}{2}^+)(\text{GeV})$&0.938&1.116&1.193 & 1.322 \\
$ M(\frac{1}{2}^-)(\text{GeV})$&1.535 &1.405&1.750&1.630        \\
$ \lambda(\text{GeV}^3) $          & 0.0255 & 0.0324 & 0.0347      & 0.0364          \\
\hline\hline
\end{tabular}
\label{table:baryon octet}
\end{table}
In our numerical calculations, related parameters of baryon octet are shown in Table~\ref{table:baryon octet} and other inputs are adopted as follows~\cite{ParticleDataGroup:2022pth, Yang:2005bv, Jido:1996zw, Nesterenko:1983ef, Hwang:1994vp, Zhao:2021sje, Wan:2021vny, Wan:2022xkx, Yang:1993bp, Jido:1996ia, Xi:2023byo, Lian:2023cgs, Du:2003ja, Ioffe:1981kw}:
\begin{align}
\label{parameter}
\langle\bar{q}q\rangle &= -(0.24 \pm 0.01)^3 \text{GeV}^3,\,  \langle\bar{s}s\rangle = (0.8 \pm 0.1)\langle\bar{q}q\rangle, \nonumber \\
m_s &= 93.4^{+8.6}_{-3.4} \,\text{MeV},\, s_{1,2}^0 = (2.0\sim 3.0)\,\text{GeV}^2\;.
\end{align}	
The masses of $ u, d $ quark are ignored. Moreover, two additional free parameters are introduced according to Eq.~(\ref{form-factors-exp}), i.e. two Borel parameters $ \tau_1^2 $ and $ \tau_2^2 $. For simplicity, we first employ the following relation of Borel parameters~\cite{Shi:2019hbf, MarquesdeCarvalho:1999bqs, Leljak:2019fqa},
\begin{align}
\frac{\tau_1^2}{\tau_2^2}=\frac{M_{1}^2-m_s^2}{M_{2}^2-m_u^2}\;,
\end{align}
where $ m_s $ and $ m_u $ denote the mass of strange and up quark, respectively.

Typically, two criteria are utilized to ascertain the values of Borel parameters. The first criterion involves the pole contribution. To examine the contribution of ground-state hadrons, it is essential for the pole contribution to dominate the spectrum. Therefore, a selection of the pole contribution larger than $ 50\% $ for the transition form factors is often made, formulated as follows:
\begin{align}
R_1^{\text{PC}}=\frac{\int_{s_1^\text{min}}^{s_1^0}d s_1\int_{s_2^\text{min}}^{s_2^0}d s_2}{\int_{s_1^\text{min}}^{\infty}d s_1\int_{s_2^\text{min}}^{s_2^0}d s_2},\quad R^{\text{PC}}_{2}=\frac{\int_{s_1^\text{min}}^{s_1^0}d s_1\int_{s_2^\text{min}}^{s_2^0}d s_2}{\int_{s_1^\text{min}}^{s_1^0}d s_1\int_{s_2^\text{min}}^{\infty}d s_2}\;.
\end{align}
These two ratios can be interpreted as the pole contributions from the initial and final baryon channel, respectively.

The second criterion is the convergence of OPE, ensuring the validity of the truncated OPE. In Eq.~(\ref{spectra-density}), nonperturbative effects up to dimension 6 are taken into account, implying that the relative contribution from the highest dimension of the condensate $ \langle \bar{q}q\rangle^2 $ needs to be below 20\%. Furthermore, as physical observables, the form factors should in principle be independent of any artificial parameters. Hence, a reliable result should be obtained within an optimal region where the form factors exhibit minimal dependence on $ \tau_{1}^2 $ and $ \tau_{2}^2 $. In practice, we adjust the threshold parameters $ s_{1,2}^0 $ by $ 0.1\, \text{GeV}^2 $ to find the acceptable range of the Borel parameters, where the form factors exhibit variations of no more than approximately 10\%.

With the above preparation, the form factors of HSDs can be numerically derived. The form factor results directly evaluated by QCDSR are shown in Table~\ref{table:numerical results}, where the uncertainties stem from the variation of the Borel parameters and the input parameters listed in Eq.~(\ref{parameter}). We have identified that the most significant errors arise from the threshold parameters $ s_{1,2}^0 $, while a recently proposed inverse problem method holds promise in mitigating the errors introduced by these threshold parameters~\cite{Li:2020ejs, Xiong:2022uwj}. The corresponding ratios are also listed in Table~\ref{table:ratios} and compared with other methods, where $ N $, $ \Lambda $, $ \Sigma $, and $ \Xi $ denote different isospin multiplet in the baryon octet.

\begin{table}[ht]
\centering
\caption{Form factor predictions at $ q^2=0 $ in QCD sum rules. The units of the Borel parameter $ \tau_2^2 $ is $ \text{GeV}^2 $.}
\resizebox{\linewidth}{!}{
\begin{tabular}{lccccccc}
\hline\hline
& $f_1(0)$ & $ \tau_2^2 $ & $f_2(0)$& $ \tau_2^2 $ & $g_1(0)$& $ \tau_2^2 $   \\
\hline
$\Lambda\rightarrow p$ & $-1.179\pm 0.075$ & $ 1.4\sim 1.8 $ &$-0.888\pm 0.074$& $ 2\sim 3 $ &$-0.843\pm 0.010$& $ 10\sim 13 $ \\
$\Sigma^{0}\rightarrow p$ & $-0.702\pm 0.042$&$ 4\sim 5 $ &$0.725\pm 0.041$& $ 7\sim 10 $&$0.229\pm 0.029$& $ 1.5\sim 2 $\\
$\Xi^{-}\rightarrow \Lambda$ & $1.216\pm 0.075$& $ 10\sim 20 $ &$0.118\pm 0.029$& $ 10\sim 20 $ &$0.328\pm 0.036$& $ 10\sim 20 $\\
$\Xi^{-}\rightarrow \Sigma^{0}$ & $0.676\pm 0.035$& $ 15\sim 20 $ &$1.394\pm 0.150$&$ 15\sim 20 $ &$0.874\pm 0.048$& $ 1.5\sim 2 $\\
\hline\hline
\end{tabular}
}
\label{table:numerical results}
\end{table}

\begin{table}[th]
\centering
\caption{Corresponding ratios of form factors.}
\begin{tabular}{lcccc}
\hline\hline
& $\Lambda\rightarrow N$ & $\Sigma\rightarrow N$ & $\Xi\rightarrow \Lambda$ & $\Xi\rightarrow \Sigma$ \\
\hline
$ f_1(0)/f_1^{SU(3)} $ \\
This work &  $0.963\pm 0.061$ & $0.993\pm 0.059$ & $ 0.993\pm 0.061$ & $0.956\pm 0.049$\\
Quark model~\cite{Donoghue:1986th}& $0.987$ & $0.987$ & $0.987$ &$0.987$ \\
Quark model~\cite{Schlumpf:1994fb} & $0.976$ & $0.975$ & $0.976$ &$0.976$ \\
$ \chi $PT~\cite{Villadoro:2006nj} & $1.027$ & $1.041$ & $1.043$ & $ 1.009 $\\
$ \chi $PT~\cite{Geng:2009ik} & $1.001^{+0.013}_{-0.010}$ & $1.087^{+0.042}_{-0.031}$ & $1.040^{+0.028}_{-0.021}$ &$1.017^{+0.022}_{-0.016}$\\
$ 1/N_c $ expansion~\cite{Flores-Mendieta:1998tfv}& $1.02\pm 0.02$ & $1.04\pm 0.02$ & $1.10\pm 0.04$ & $1.12\pm 0.05$\\
lattice QCD~\cite{Sasaki:2012ne} &  & $0.957\pm 0.01$ &  &$0.976\pm 0.005$\\
\hline
$ g_1(0)/f_1(0) $ \\
This work &  $0.708\pm 0.047$ & $-0.327\pm 0.046$ & $ 0.271\pm 0.045$ & $1.293\pm 0.100$\\
Cabibbo model~\cite{Cabibbo:2003cu} &  $0.731$ & $-0.341$ & $ 0.195$ & $1.267$\\
Quark model~\cite{Faessler:2008ix} &  $0.724$ & $-0.260$ & $ 0.265$ & $1.20$\\
Soliton  model~\cite{Yang:2015era} &  $0.718\pm 0.003$ & $-0.340\pm 0.003$ & $ 0.250\pm 0.002$ & $1.210\pm 0.005$\\
Soliton  model~\cite{Ledwig:2008ku} &  $0.68$ & $-0.27$ & $ 0.21 $ & $1.16$\\
$ 1/N_c $ expansion~\cite{Flores-Mendieta:1998tfv} &  $0.73$ & $-0.34$ & $ 0.22$ & $1.03$\\
lattice QCD~\cite{Guadagnoli:2006gj, Sasaki:2008ha} &   & $-0.287\pm 0.052$ &  & $1.248\pm 0.029$\\
Exp~\cite{ParticleDataGroup:2022pth}&  $0.718\pm 0.015$ & $-0.340\pm 0.017$ & $ 0.25\pm 0.05$ & $1.22\pm 0.05$\\
\hline
$ f_2(0)/f_1(0) $ \\
This work &  $0.752\pm 0.074$ & $-1.042\pm 0.090$ & $ 0.118\pm 0.032$ & $1.957\pm 0.255$\\
Cabibbo model~\cite{Cabibbo:2003cu} &  $1.066$ & $-1.292$ & $ 0.085$ & $2.609$\\
Quark model~\cite{Faessler:2008ix} &  $1$ & $-0.962$ & $ 0.129$ & $2.402$\\
Soliton  model~\cite{Yang:2015era} &  $0.637\pm 0.041$ & $-0.709\pm 0.036$ & $ -0.069\pm 0.027$ & $1.143\pm 0.061$\\
Soliton  model~\cite{Ledwig:2008ku} &  $0.71$ & $-0.96$ & $ -0.02$ & $2.02$\\
$ 1/N_c $ expansion~\cite{Flores-Mendieta:1998tfv} &  $0.90$ & $-1.02$ & $ -0.06$ & $1.85$\\
lattice QCD~\cite{Sasaki:2008ha} &   & $-1.52\pm 0.81$ &  & \\
Exp~\cite{ParticleDataGroup:2022pth}& $ 1.32\pm 0.81 $~\cite{Bristol-Geneva-Heidelberg-Orsay-Rutherford-Strasbourg:1983jzt} & $-0.97\pm 0.14$ & $ -0.24\pm 0.25 $~\cite{Bristol-Geneva-Heidelberg-Orsay-Rutherford-Strasbourg:1983jzt} & $2.0\pm 0.9$\\
\hline\hline
\end{tabular}
\label{table:ratios}
\end{table}

The ratio $ f_1(0)/f_1^{SU(3)} $ describes the flavor $ SU(3) $ symmetry breaking effects of $ f_1 $, where the values of $ f_1^{SU(3)} $ are taken from Cabibbo theory~\cite{Cabibbo:2003cu}. Early results of $ f_1(0)/f_1^{SU(3)} $ from two quark models~\cite{Donoghue:1986th, Schlumpf:1994fb} exhibit remarkable similarity across the four decay channels, displaying a consistent negative sign for the $ SU(3) $ corrections of $ f_1 $. In contrast, analyses employing chiral perturbation theory~\cite{Villadoro:2006nj, Geng:2009ik} and $ 1/N_c $ expansion~\cite{Flores-Mendieta:1998tfv} generally favor positive signs. Recent lattice QCD results~\cite{Sasaki:2012ne} support a negative correction for $ f_1 $, aligning with our central values. Additionally, we note that the $ f_1(0)/f_1^{SU(3)} $ value for $\Xi\rightarrow \Sigma$ is the smallest among the four decay channels, suggesting relatively larger breaking effects of $ f_1 $ in the $\Xi\rightarrow \Sigma$ mode. This tendency is also in agreement with $ 1/N_c $ expansion results.

The ratio $ g_1(0)/f_1(0) $ is characterized by $ F $ and $ D $ in Eq.~(\ref{g10}), where a fit of these two constants using old experimental data yields $ F\approx 0.463 $ and $ D\approx 0.804 $~\cite{Cabibbo:2003cu}. In Table~\ref{table:ratios}, it can be observed that the results of $ g_1(0)/f_1(0) $ from various theoretical methods, along with experimental data, generally maintain overall consistency within the margin of error. It is noteworthy that the $\Xi\rightarrow \Sigma$ mode holds a particularly unique status as it is identical to the well measured neutron beta decay under the exchange of the down quark with the strange quark~\cite{Sasaki:2008ha, KTeV:2001djr}. In this case, only the CKM matrix elements become important to distinguish these decay amplitudes~\cite{Ledwig:2008ku}, and the flavor $SU(3)$ breaking effects can be easily exposed through a comparison with results of neutron beta decay~\cite{Sasaki:2008ha}. The experimental result of $ g_1(0)/f_1(0) $ for $\Xi\rightarrow \Sigma$ mode given in Table~\ref{table:ratios} represents the world average taken from the $\Xi^0\rightarrow \Sigma^+$ process in Particle Data Group~\cite{ParticleDataGroup:2022pth}, while our central value is more closely aligned with the results announced by KTeV collaboration with $ g_1(0)/f_1(0) = 1.32_{-0.17}^{+0.21}\pm 0.05 $~\cite{KTeV:2001djr}. The precise experimental value of $ g_1(0)/f_1(0) $ for neutron beta decay is $ g_1(0)/f_1(0) = 1.2754\pm 0.013 $~\cite{ParticleDataGroup:2022pth}, and it slightly deviates from the current data as well as our calculated results of the $\Xi^0\rightarrow \Sigma^+$ process. Furthermore, the isospin partner process $\Xi^-\rightarrow \Sigma^0$ was measured in Ref.~\cite{Bristol-Geneva-Heidelberg-Orsay-Rutherford-Strasbourg:1983jzt}, and they reported a ratio of $ g_1(0)/f_1(0) = 1.25^{+0.14}_{-0.16} $. If the form factors for the $\Xi^-\rightarrow \Sigma^0$ decay can be measured with greater precision, it may provide insights into isospin symmetry breaking.

As for the ratio $ f_2(0)/f_1(0) $, considering the relatively large uncertainties of the experimental data presented in Table~\ref{table:ratios}, it is challenging to make a direct comparison between the theoretical results and the existing data. Notice, results from different theoretical approaches still exhibit significant disparities, further investigations are needed to determine the precise value of the weak magnetism form factor $ f_2 $.

\begin{figure}[ht]
\centering
\includegraphics[width=7.3cm]{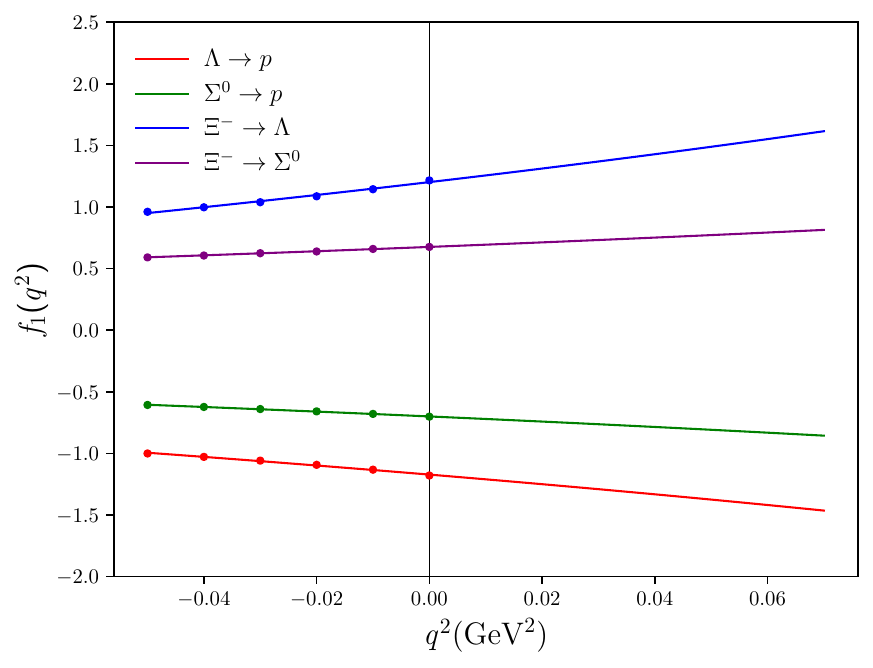}
\includegraphics[width=7.3cm]{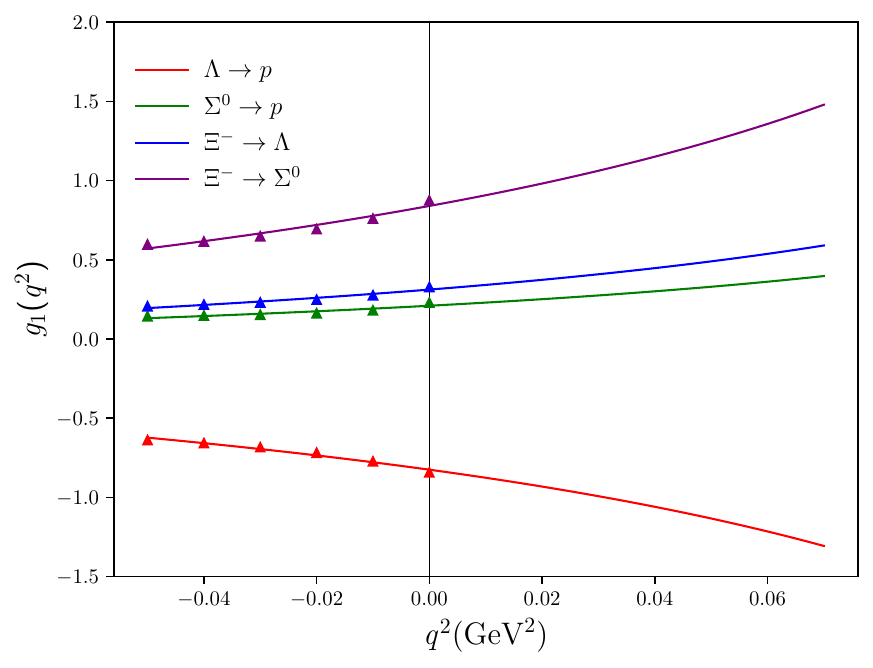}
\caption{The $ q^2 $ dependence of form factors with the central value of fitting parameters listed in Table~\ref{table:fit}. Moreover, the symbol point, as well as the triangle, represent the fitted points for each of the form factors.}
\label{fig:ft}
\end{figure}

To obtain the $ q^2 $ dependence of the form factors, we calculate them at a small space-like interval $ q^2 \in [-0.05,0]$ $\text{GeV}^2  $. Then we fit the data by employing the $ z $-series parameterization~\cite{Bourrely:2008za}, which is widely used in the fitting of form factors~\cite{Xing:2021enr, Huang:2022lfr, Han:2023pgf, Zhao:2020mod, Khodjamirian:2011jp, Leljak:2019fqa}. The expression of $ z $-series parameterization is as follows:
\begin{align}
\label{BCL}
f_i(q^2)&=\frac{f_i(0)}{1-q^2/(m_{pole})^2}\Bigl\{1+a_1(z(q^2,t_0)-z(0,t_0))\Bigr\},\nonumber\\[10pt]
z(q^2,t_0)&=\frac{\sqrt{t_+-q^2}-\sqrt{t_+-t_0}}{\sqrt{t_+-q^2}+\sqrt{t_+-t_0}}\;.
\end{align}
Here, $ a_1 $ is a fitting parameter, and $ f_i(0) $ represents the value of form factors at $ q^2=0 $, which is also treated as a fitted parameter. The variables $ t_{\pm}= (M_{1}\pm M_2)^2$, and $ t_0=t_+-\sqrt{t_+-t_-} \sqrt{t_+-t_{min}}$. In the numerical analysis, we choose $ t_{min}=-0.05 \,\text{GeV}^2 $ and $ m_{pole}=m_{K^*}=0.892\,\text{GeV} $ for $ f_1 $ and $ m_{pole}=m_{K_1}=1.27\,\text{GeV} $ for $ g_1 $~\cite{ParticleDataGroup:2022pth}. The nonlinear least squares ($ \chi^2 $) fitting method are employed in our analysis. The fitting results are listed in Table~\ref{table:fit}, where the values of $ f_1(0) $ and $ g_1(0) $ obtained from the fitting procedure are in agreement with our directly calculated results provided in Table~\ref{table:numerical results}. The $ q^2 $ dependence of form factors for all HSDs process are shown in Fig.~\ref{fig:ft}. It should be mentioned that the $ q^2 $ dependence of $ f_2 $ can be neglected since its contribution to the decay amplitude is already $ O(q) $~\cite{Cabibbo:2003cu}. Besides, the results of $ \Sigma^- \rightarrow n $ and $ \Xi^{0}\rightarrow \Sigma^{+} $ can be derived using isospin relation~\cite{Yang:2015era, Ledwig:2008ku}:
\begin{align}
(\Sigma^- \rightarrow n)=\sqrt{2}\,(\Sigma^0 \rightarrow p),\,\,\,(\Xi^0\to\Sigma^+)=\sqrt{2}\,(\Xi^-\to\Sigma^0)\;.
\end{align}

\begin{table}[ht]
\centering
\caption{Summary of the fitted parameters for the form factors.}
\begin{tabular}{lcccc}
\hline\hline
& $f_1(0)$ & $a_1$ & $g_1(0)$ & $a_1$  \\
\hline
$\Lambda\rightarrow p$ & $-1.172\pm 0.075$& $-33.31\pm 1.25$ &$-0.820\pm 0.007$& $-74.98\pm 1.48$\\
$\Sigma^{0}\rightarrow p$ & $-0.700\pm 0.041$& $-29.64\pm 2.81$ &$0.208\pm 0.027$& $-128.27\pm 7.32$ \\
$\Xi^{-}\rightarrow \Lambda$ & $1.202\pm 0.072$& $-76.14\pm 11.51$ &$0.309\pm 0.034$& $-169.04\pm 2.19$ \\
$\Xi^{-}\rightarrow \Sigma^{0}$ & $0.676\pm 0.027$& $-35.40\pm 31.72$ &$0.833\pm 0.033$& $-150.38\pm 25.65$ \\
\hline\hline
\end{tabular}
\label{table:fit}
\end{table}

Equipped with the $ q^2 $ dependence of the form factors, the branching fractions of HSDs can be systematically analyzed. For convenience, we introduce the following helicity amplitudes~\cite{Shi:2019hbf,Huang:2022lfr,  Wang:2019alu}:
\begin{align}
& H_{\frac{1}{2}, 0}^V= \frac{\sqrt{Q_{-}}}{\sqrt{q^2}}\big(M_{+} f_1(q^2)-\frac{q^2}{M_{1}} f_2(q^2)\big), \quad H_{\frac{1}{2}, 0}^A= \frac{\sqrt{Q_{+}}}{\sqrt{q^2}}\big(M_{-} g_1(q^2)+\frac{q^2}{M_{1}} g_2(q^2)\big)\ , \nonumber\\
& H_{\frac{1}{2}, 1}^V= \sqrt{2 Q_{-}}\big(-f_1(q^2)+\frac{M_{+}}{M_{1}} f_2(q^2)\big), \quad H_{\frac{1}{2}, 1}^A= \sqrt{2 Q_{+}}\big(-g_1(q^2)-\frac{M_{-}}{M_{1}} g_2(q^2)\big)\ , \nonumber\\
& H_{\frac{1}{2}, t}^V= \frac{\sqrt{Q_{+}}}{\sqrt{q^2}}\big(M_{-} f_1(q^2)+\frac{q^2}{M_{1}} f_3(q^2)\big), \quad H_{\frac{1}{2}, t}^A= \frac{\sqrt{Q_{-}}}{\sqrt{q^2}}\big(M_{+} g_1(q^2)-\frac{q^2}{M_{1}} g_3(q^2)\big)\;.
\end{align}
Here, $ H_{\lambda^\prime, \lambda_W}^{V(A)} $ is the helicity amplitudes for weak transitions induced by vector and axial-vector currents, where $ \lambda^\prime $ and $ \lambda_W $ are the helicity of the final state baryon and the virtual $ W $ boson, respectively. $ Q_{\pm} $ is defined as $ Q_{\pm} = M_{\pm}^2-q^2 $ and $ M_{\pm}=M_{1}\pm M_2 $. With the above helicity amplitudes, the differential distribution of HSDs process $ B_1\rightarrow B_2\ell\nu_\ell $ can be written as:
\begin{align}
\label{differential}
&\frac{d \Gamma\left(B_1\rightarrow B_2\ell\nu_\ell\right)}{d q^2}=\frac{G_F^2\left|V_{u s}\right|^2 q^2\sqrt{Q_+Q_-}}{384 \,\pi^3 \,M_{1}^3}(1-\frac{m_\ell^2}{q^2})^2 H_{\text{tot}}\;,
\end{align}
where $ m_{\ell} $ represents the lepton mass ($ \ell=e,\mu $) and $ H_{\text{tot}} $ is defined as
\begin{align}
H_{\text{tot}} &= \big(1+\frac{m_\ell^2}{2 q^2}\big)\big(H_{\frac{1}{2},1}^2+H_{-\frac{1}{2},-1}^2+H_{\frac{1}{2},0}^2+H_{-\frac{1}{2},0}^2\big)+\frac{3\,m_\ell^2}{2 q^2}\big(H_{\frac{1}{2},t}^2+H_{-\frac{1}{2},t}^2\big)\;.
\label{Htot}
\end{align}
Here, the total helicity amplitudes, $ H_{\lambda^\prime, \lambda_W} $, can be derived using the following relations:
\begin{align}
H_{-\lambda^\prime,-\lambda_W}^V&=H_{\lambda^\prime, \lambda_W}^V, \quad  \quad H_{-\lambda^\prime,-\lambda_W}^A=-H_{\lambda^\prime, \lambda_W}^A,\nonumber\\[5pt]
H_{\lambda^\prime, \lambda_W}&=H_{\lambda^\prime, \lambda_W}^V-H_{\lambda^\prime, \lambda_W}^A\;.
\end{align}
From the definition of $ H_{\text{tot}} $ in Eq.~(\ref{Htot}), one may notice that the contribution to the differential decay width from $ f_3(q^2) $ and $ g_3(q^2) $ is presented in the term $ H_{\frac{1}{2},t}^2 $ and $ H_{-\frac{1}{2},t}^2 $, which is evidently suppressed by a factor of $ m_\ell^2 $. Therefore, we opt to neglect the effect of $ f_3(q^2) $ and $ g_3(q^2) $ in Eq.(\ref{matrix}). To compute the numerical results for the differential decay width, we utilize the following input parameters relevant to the decay analysis~\cite{ParticleDataGroup:2022pth}:
\begin{align}
&m_e=0.511\,\text{MeV},\quad m_\mu=0.106\,\text{GeV},\quad G_F=1.166\times 10^{-5}\, \text{GeV}^{-2},\quad |V_{us}|=0.2243,\nonumber\\
&\tau_{\Lambda}=(2.632\pm 0.020)\times 10^{-10} s, \quad \tau_{\Sigma^-}=(1.479\pm 0.011)\times 10^{-10} s,\quad \tau_{\Sigma^0}=(74\pm 7)\times 10^{-21} s,\nonumber\\
&\tau_{\Xi^0}=(2.90\pm 0.09)\times 10^{-10} s,\quad \tau_{\Xi^-}=(1.639\pm 0.015)\times 10^{-10} s\;.
\end{align}

\begin{figure}[th]
\centering
\includegraphics[width=7.3cm]{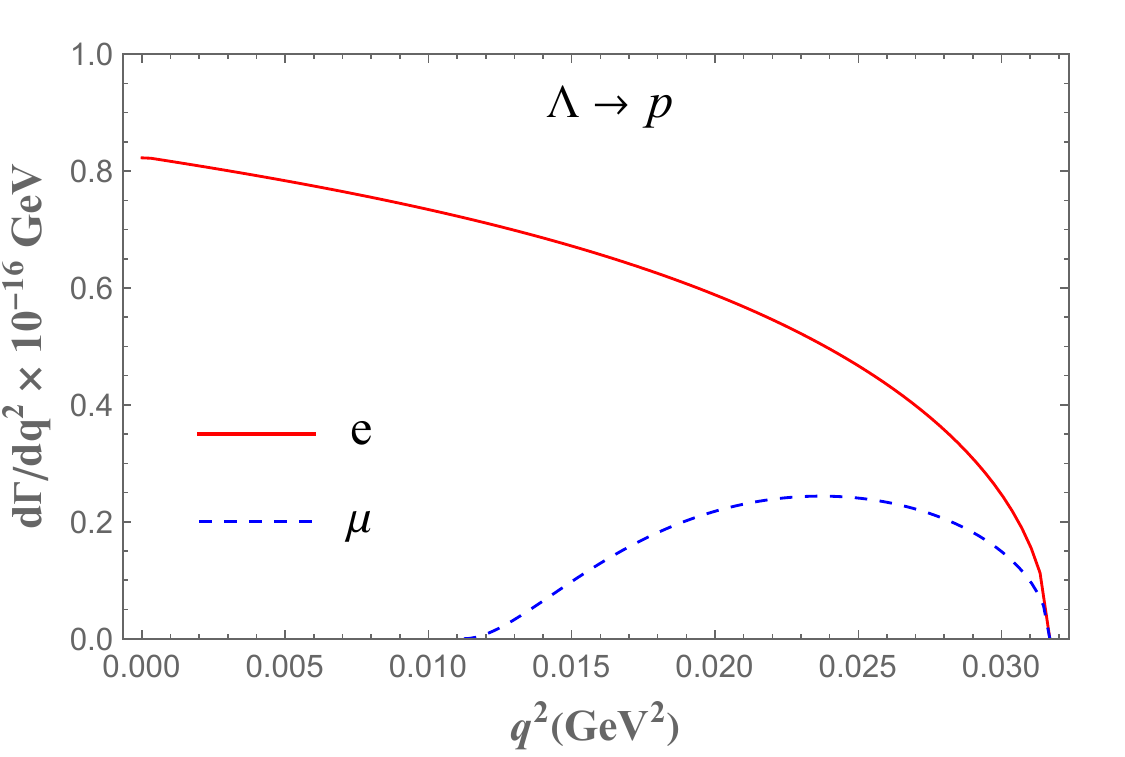}
\includegraphics[width=7.3cm]{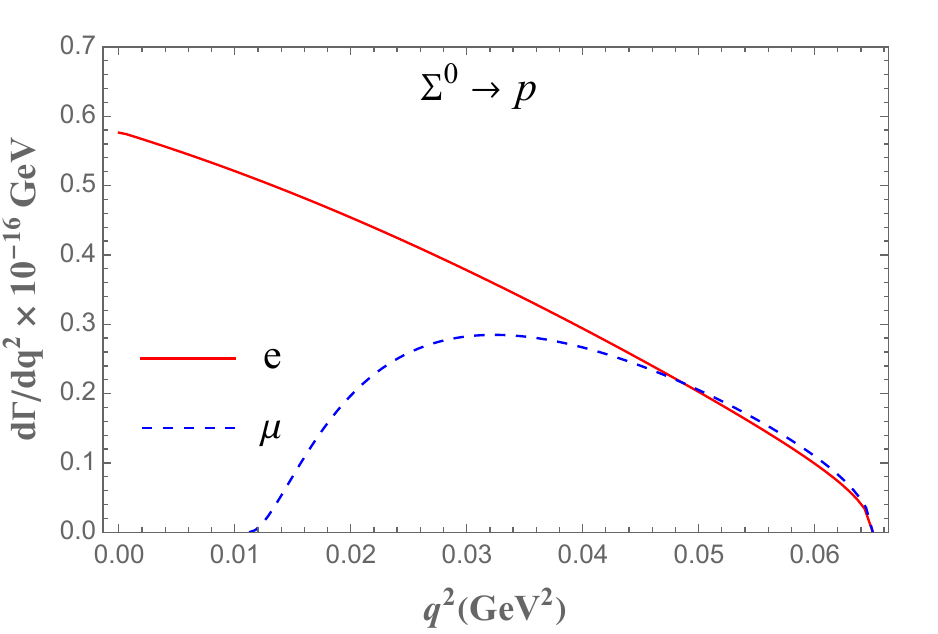}
\includegraphics[width=7.3cm]{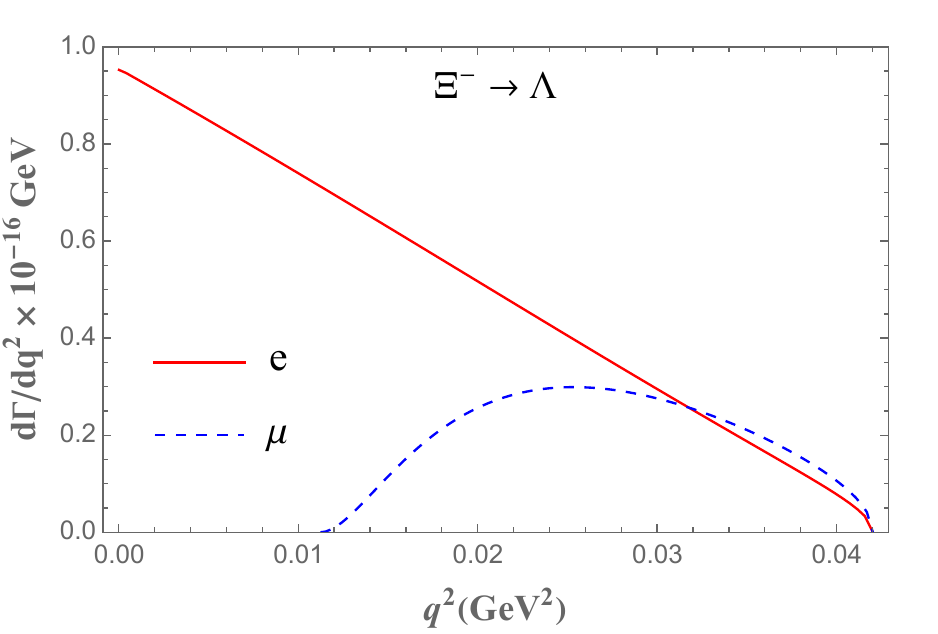}
\includegraphics[width=7.3cm]{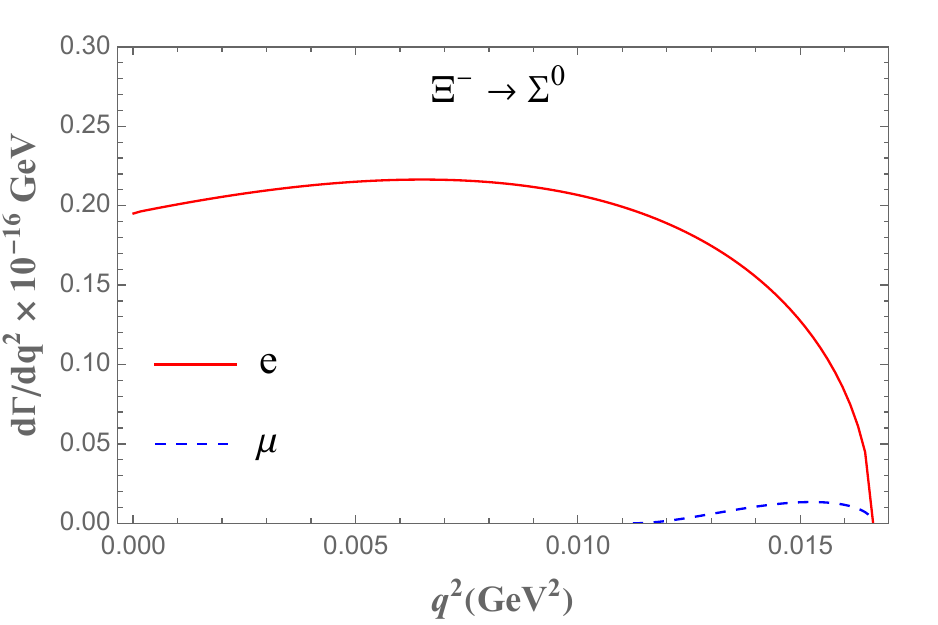}
\caption{The $ q^2 $ dependence of the differential decay width for HSDs process $ B_1\rightarrow B_2 \ell\nu_\ell $. The red solid line denotes $ \ell=e^- $, while the blue dashed line denotes $ \ell=\mu^- $.}
\label{fig:width}
\end{figure}

Here, the mean lifetime of hyperons, noted as $ \tau $, is introduced to derive the branching fractions. To determine the total decay width, we must first calculate the differential decay width. Eq.~(\ref{differential}) enables us to compute the differential decay width for electronic and muonic mode, and the corresponding $ q^2 $ dependence of the differential decay width are plotted in Fig.~\ref{fig:width}. These are merely theoretical predictions of the $ q^2 $ dependence for the differential decay width in HSDs, and as of now, there is no experimental measurement available due to the extremely limited kinematic regions.
The global behavior of the curve closely resembles that in Ref.~\cite{Migura:2006en} using a relativistic quark model, profoundly impacted by the factor $ q^2\sqrt{Q_+Q_-} $. Moreover, the differential decay width of electronic mode does not vanish at $ q^2 = 0 $ since the limit behavior of the helicity amplitudes, while it becomes zero at $ q^2 = m_\mu^2 $ for muonic mode. As for the area around $ q^2 = m_e^2 $, since the electron mass squared $ m_e^2\approx 0.25\times 10^{-6}\text{GeV}^2 $ is tiny compared to the typical range of $ q^2\sim10^{-2}\text{ GeV}^2 $, there will be an indistinguishable very sharp fall-off for the electronic mode.

Integrating over the $q^2$ of those curves in Fig.~\ref{fig:width} we obtain the total decay width, ready for the determination of the decay branching fractions. The normal branching fraction results for HSDs are presented in the second column of Table~\ref{table:br}. By disregarding the $ q^2 $ dependence of the form factors and employing the directly calculated values at $ q^2 = 0 $ from Table~\ref{table:numerical results}, we have included the corresponding results for the branching fractions in the third column of Table~\ref{table:br}. In Table~\ref{table:br}, it is evident that the inclusion of $q^2$ dependence of the form factors has a notable impact on the branching fractions, and enhances the consistency with the experimental data. Specifically, the $q^2$ dependence of the form factors results in a contribution of approximately 10\% to 20\% to the branching fractions. Additionally, the influence of $q^2$ dependence of the form factors is typically more pronounced in the muonic mode compared to the electronic mode. These characters agree with the conclusions in Ref.~\cite{Faessler:2008ix} exploiting a Lorentz covariant quark model approach.

Note that the results of $ \Sigma^0\rightarrow p $ process present in the fourth column of Table~\ref{table:br} are provided in Ref.~\cite{Wang:2019alu} using $ SU(3) $ flavor symmetry due to the lack of experimental data of this transition. It should be emphasised that the experimental data extracted from the Particle Data Group mostly has a long history with large uncertainties, and many of them are not direct measurements but rely on corresponding two-body hadronic decays~\cite{ParticleDataGroup:2022pth}. Besides, the number of events of the muonic mode is scarce, especially for the transitions $ \Xi^{-}\rightarrow \Lambda\mu^-\bar{\nu}_\mu $ and $ \Xi^{-}\rightarrow \Sigma^0\mu^-\bar{\nu}_\mu $. The former exhibits significant uncertainties, and the latter only yields a upper limit. Therefore, more precise measurements for HSDs are highly necessary.

\begin{table}[ht]
\centering
\caption{Branching fractions for HSDs process $ B_1\rightarrow B_2\ell\nu_\ell $ in QCD sum rules. The superscript 0 in the third column represents the results without $ q^2 $ dependence of the form factors.}
\begin{tabular}{lccccccc}
\hline\hline
&& & QCDSR & & QCDSR$ ^0 $ & & Exp~\cite{ParticleDataGroup:2022pth}\\
\hline
$\mathcal{B}(\Lambda\rightarrow p e^- \bar{\nu}_e)\times 10^{-4}$& &  & $7.72\pm 0.64$&&$7.12\pm 0.70$ & &$8.34\pm 0.14$  \\
$\mathcal{B}(\Lambda\rightarrow p\mu^-\bar{\nu}_\mu)\times 10^{-4}$& &  & $1.35\pm 0.11$&  & $1.15\pm 0.11$ & &$1.51\pm 0.19$  \\
$\mathcal{B}(\Sigma^{-}\rightarrow n e^- \bar{\nu}_e)\times 10^{-3}$& & & $1.00\pm 0.18$& & $0.87\pm 0.14$ & &$1.017\pm 0.034$ \\
$\mathcal{B}(\Sigma^{-}\rightarrow n\mu^-\bar{\nu}_\mu)\times 10^{-4}$& & & $4.73\pm 0.88$& & $3.74\pm 0.62$ & &$4.5\pm 0.4$ \\
$\mathcal{B}(\Sigma^{0}\rightarrow p e^- \bar{\nu}_e)\times 10^{-13}$& & &$2.50\pm 0.70$ && $2.18\pm 0.58$ && $2.46\pm 0.32$~\cite{Wang:2019alu}\\
$\mathcal{B}(\Sigma^{0}\rightarrow p\mu^-\bar{\nu}_\mu)\times 10^{-13}$& & &$1.18\pm 0.34$ && $0.94\pm 0.25$ &&$1.13\pm 0.15$~\cite{Wang:2019alu} \\
$\mathcal{B}(\Xi^{-}\rightarrow \Lambda e^- \bar{\nu}_e)\times 10^{-4}$& & & $5.17\pm 0.73$ & & $4.67\pm 0.66$ & & $5.63\pm 0.31$ \\
$\mathcal{B}(\Xi^{-}\rightarrow \Lambda\mu^-\bar{\nu}_\mu)\times 10^{-4}$& & & $1.54\pm 0.23$ & & $1.24\pm 0.18$ & & $3.5^{+3.5}_{-2.2}$ \\
$\mathcal{B}(\Xi^{0}\rightarrow \Sigma^{+} e^- \bar{\nu}_e)\times 10^{-4}$& & &$2.41\pm 0.28$ & & $2.39\pm 0.32$ & & $2.52\pm 0.08$\\
$\mathcal{B}(\Xi^{0}\rightarrow \Sigma^{+}\mu^-\bar{\nu}_\mu)\times 10^{-6}$& & &$2.07\pm 0.27$ && $1.91\pm 0.25$& & $2.33\pm 0.35$\\
$\mathcal{B}(\Xi^{-}\rightarrow \Sigma^{0} e^- \bar{\nu}_e)\times 10^{-5}$& &  & $7.80\pm 0.77$ & & $7.72\pm 0.86$ & & $8.7\pm 1.7$\\
$\mathcal{B}(\Xi^{-}\rightarrow \Sigma^{0}\mu^-\bar{\nu}_\mu)\times 10^{-6}$& &  & $1.06\pm 0.12$ & & $0.97\pm 0.11$ & & $ <800$\\
\hline\hline
\end{tabular}
\label{table:br}
\end{table}

In conclusion, we focus on the non-standard form factor introduced by the tensor interaction~\cite{Weinberg:1958ut, Goudzovski:2022vbt}:
\begin{align}
\label{tensor}
\bra{B_2(q_2)}\bar{u}\,\sigma_{\mu \nu}\,s\ket{B_1(q_1)}\simeq f_T(q^2)\,\bar{u}_{2}(q_2)\,\sigma_{\mu \nu} \,u_1(q_1)\;,
\end{align}
where $ f_T $ represents the tensor form factor. The matrix element in Eq.~(\ref{tensor}) is essential for calculating the production of the massless dark photon induced by the magnetic dipole moment (MDM)~\cite{Goudzovski:2022vbt}. Furthermore, other terms to the matrix element of the tensor current are neglected since they are kinematically suppressed~\cite{Chang:2014iba}. By applying the same procedure, the values of the tensor form factors $ f_T(0) $ as well as the ratio $ f_T(0)/f_1(0) $ can be determined, which are presented in Table~\ref{table:tensor}. It can be observed that the values of $ f_T(0)/f_1(0) $ in each channel align consistently with Ref.~\cite{Chang:2014iba}, which are very close to the experimental data of $ g_1(0)/f_1(0) $.
\begin{table}[th]
\centering
\caption{Determination of $ f_T(0) $ and $ f_T(0)/f_1(0) $ in QCDSR.}
\begin{tabular}{lcccc}
\hline\hline
& $\Lambda\rightarrow p$ & $\Sigma^0\rightarrow p$ & $\Xi^-\rightarrow \Lambda$ & $\Xi^-\rightarrow \Sigma^0$ \\
\hline
$ f_T(0) $ &  $-0.777\pm 0.060$ & $0.195\pm 0.018$ & $ 0.349\pm 0.037$ & $0.861\pm 0.053$\\
$ f_T(0)/f_1(0) $ &  $0.663\pm 0.070$ & $-0.284\pm 0.046$ & $ 0.288\pm 0.051$ & $1.278\pm 0.112$\\
$ f_T(0)/f_1(0) $~\cite{Chang:2014iba} &  $0.72$ & $-0.28$ & $ 0.22$ & $1.22$\\
$ g_1(0)/f_1(0) $~\cite{ParticleDataGroup:2022pth}&  $0.718\pm 0.015$ & $-0.340\pm 0.017$ & $ 0.25\pm 0.05$ & $1.22\pm 0.05$\\
\hline\hline
\end{tabular}
\label{table:tensor}
\end{table}

\section{Conclusions}

In this work, we utilize QCD sum rules to investigate the hyperon semileptonic decays. Particularly, we compute the vector form factor $ f_1 $, the weak magnetism form factor $ f_2 $, and the axial-vector form factor $ g_1 $ for all the pertinent hyperon semileptonic decays. The Cutkosky cutting rules is employed to derive the analytic results of the relevant form factors. We neglect $ f_3 $ and $ g_3 $ due to their substantially suppressed contributions to the decays and also neglect the second-class form factor $ g_2 $ which is often assumed to be small.

We first derive the numerical value for the form factors at $ q^2=0 $, and then discuss the corresponding ratios of these form factors. Our findings regarding $ f_1(0)/f_1^{SU(3)} $ exhibit a consistent trend with the quark model~\cite{Donoghue:1986th, Schlumpf:1994fb} and lattice QCD methods~\cite{Sasaki:2012ne}, and there are disparities compared to the results derived from chiral perturbation theory and $ 1/N_c $ expansion~\cite{Villadoro:2006nj, Geng:2009ik, Flores-Mendieta:1998tfv}. For the value of $ g_1(0)/f_1(0) $, different theoretical methods as well as the experimental data yield approximately the same results. The variation of $ g_1(0)/f_1(0) $ between $ \Xi^0\rightarrow \Sigma^+ \ell\nu_\ell $ and neutron beta decay indicates potential flavor $ SU(3) $ symmetry breaking in HSDs. The last ratio, $ f_2(0)/f_1(0) $, exhibits substantial variations among different theoretical predictions and experimental data. Therefore, additional efforts are required to thoroughly investigate the weak magnetism form factor $ f_2 $.

Based on the $ q^2 $ dependence of the form factors, we present the analysis of differential decay width and branching fractions. Our numerical results for HSDs could provide useful information to reliably extract the value of the CKM matrix element $ V_{us} $. Finally, we conduct explorations on the tensor form factor $ f_T $ related to the new physics beyond the standard model, and find that there is still room for further refinement in the QCD sum rule method. As previously discussed, the sensitivity of the results to threshold parameters may be alleviated through the application of inverse problem methods. Besides, error reduction could be achieved through the implementation of radiative corrections.

\vspace{0.5cm}
{\bf Acknowledgments}

The authors acknowledge the meaningful discussion with Q.M. Feng. This work was supported in part by the National Key Research and Development Program of China under Contracts No. 2020YFA0406400, by the National Natural Science Foundation of China(NSFC) under the Grants 11975236 and 12235008.


\end{document}